# 90 degree polarization rotator using a bilayered chiral metamaterial with giant optical activity

Yuqian Ye[1] and Sailing He[1,2,*]

[1]*Centre for Optical and Electromagnetic Research, State Key Laboratory of Modern Optical Instrumentations, Zijingang Campus, Zhejiang University, China*

[2]*Division of Electromagnetic Engineering, School of Electrical Engineering, Royal Institute of Technology, S-100 44 Stockholm, Sweden*

A bilayered chiral metamaterial (CMM) is proposed to realize a 90 degree polarization rotator, whose giant optical activity is due to the transverse magnetic dipole coupling among the metallic wire pairs of enantiomeric patterns. By transmission through this thin bilayered structure of less than $\lambda/30$ thick, a linearly polarized wave is converted to its cross polarization with a resonant polarization conversion efficiency (PCE) of over 90%. Meanwhile, the axial ratio of the transmitted wave is better than 40 dB. It is demonstrated that the chirality in the propagation direction makes this efficient cross-polarization conversion possible. The transversely isotropic property of this polarization rotator is also experimentally verified. The optical activity of the present structure is about $2700°/\lambda$, which is the largest optical activity that can be found in literature.

Metamaterials [1,2], for their striking abilities of creating an effective medium with controllable permittivity and permeability, have attracted great attraction during the last decade. Many unprecedented electromagnetic properties and functionalities unattainable from naturally occurring materials have been achieved with metamaterials, such as negative index materials [3], frequency selective cloaking [4], super lenses [5] and electromagnetic wave tunneling of zero-index material [6]. All these research efforts have been focused on controlling the propagation of electromagnetic waves by designing some resonant metamaterial elements.

Recently, metamaterials have also been used to manipulate the polarization states of electromagnetic waves [7,8]. A near-complete cross polarization conversion (CPC) has been achieved with anisotropic metamaterials due to the reflection (transmission) phase difference in the two orthogonal polarizations. However, CPC obtained through these anisotropic metamaterials is sensitive to the polarization azimuth of the incident wave. An alternative approach to the polarization conversion is by using chiral metamaterials (CMM) [9-11]. Optical activity of commonly used bilayered CMM is due to the electric dipole coupling between the mutually twisted patterns [10,11]. It is difficult to realize a polarization rotation of up to 90˚ with a high efficiency of CPC in such structures. Another chiral metamaterial based on some longitudinal magnetic dipole coupling [12,13] has been proposed to get strong rotatory strength. However, such a CMM is not transversely isotropic. Moreover, the polarization rotation in this chiral structure is accompanied by the change of polarization ellipticity. In this letter, we propose a new bilayered CMM composed of some enantiomeric patterns with four-fold rotational symmetry. Through the transverse magnetic dipoles coupling in two orthogonal directions, a 90° polarization rotation with high polarization conversion efficiency (PCE) is obtained, which is independent on the incident polarization azimuth.

Fig. 1(a) shows the schematic diagram of one unit cell of the present bilayered CMM. The transverse dimension of the unit cell is 14 mm × 14 mm, while the whole sample [see Fig. 1(b)] is composed of 20 × 20 unit cells. Taconic TLY-5 with a dielectric constant of 2.2 + 0.0009i is chosen as a substrate whose thickness $t$ is 0.762 mm. The circuit board was coated on both sides with 19 nm-thick layers of copper using conventional photolithography techniques. As shown in Fig. 1(a), the chiral pattern on the upper side consists of four identical cut-wires which form a square. The length $l$ and width $w$ of the cut-wires are 9.72 mm and 2 mm, respectively. The air gap $g$ [see Fig. 1(a)] between two adjacent cut-wires is 0.28 mm. The pattern on the bottom side [see Fig.1 (c)] is in an enantiomeric form of the upper one. Obviously, this bilayered metamaterial structure still has the four-fold rotational symmetry, while the mirror symmetry in z direction is broken. Fig. 1(c) shows the schematic picture of the experimental setup. Transmission measurement is performed by using a network analyzer (Agilent 8722ES) and a pair of horn antennas (HD-100HA20), which serve as the source and detector. The horns are positioned 50 cm away from each other and the slab sample is supported by a desk at the center. Both horns are coupled to linearly polarized waves. In our experiment, the incident wave is polarized in x-direction [see

Fig. 1(c)] and normally impinged on the sample. All the measurements were performed in an anechoic chamber.

The measured transmission of electric field for the bilayered CMM is shown by the dotted line in Fig. 2(a). The blue dotted line represents the co-polarized transmission which is defined as $|t_{xx}| = |E_x^{out}|/|E_x^{in}|$, relating the incident ($E_x^{in}$) and transmitted ($E_x^{out}$) waves. While the red one is the cross-polarized transmission defined as $|t_{yx}| = |E_y^{out}|/|E_x^{in}|$. From Fig. 2(a), one sees that $|t_{xx}|$ is less than -5 dB in the frequency range we considered and greatly suppressed to a level below -20 dB near 12 GHz. Meanwhile, an obvious resonant peak in $|t_{yx}|$ can be observed with the maximum amplitude higher than -0.2 dB around 11.4 GHz. This indicates PCE (defined as the percentage of the power transfer from the input polarized mode to the output orthogonally polarized mode) of over 90% is achieved through the strong rotatory strength in this bilayered CMM. At this resonant frequency, the co-polarized component of the transmitted wave is rather small (about -15 dB), i.e., the transmitted wave is almost linearly polarized in y-direction. Therefore, we can deduce that a 90° polarization rotator with high PCE is obtained. The above measurement results are confirmed by the numerical results calculated by using a finite-integration time domain algorithm (CST Microwave Studio). The simulated transmission of electric field is shown by the solid lines in Fig. 2(a), which agrees well with the experimental results despite the fabrication error and random error in measurements.

To understand better the dependence of this cross-polarized conversion on the chirality of the present structure (with four-fold rotational symmetry), we use the Jones calculus to describe the wave transmission process (assuming the transmitted wave is fully polarized, such as linearly polarized or elliptically polarized). Let $T = \begin{Bmatrix} t_{xx} & t_{xy} \\ t_{yx} & t_{yy} \end{Bmatrix}$ denotes the transmission matrix for this bilayered structure. Then the wave transmission process mentioned above (with the incident wave polarized along x direction) can be described as

$$\begin{Bmatrix} t_{xx} & t_{xy} \\ t_{yx} & t_{yy} \end{Bmatrix} \times \begin{pmatrix} E_x^{in} \\ 0 \end{pmatrix} = \begin{pmatrix} E_x^{out} \\ E_y^{out} \end{pmatrix}. \tag{1}$$

Since the present structure is of four-fold rotational symmetry, we have $\mathfrak{R}_{90} T \mathfrak{R}_{90}^{-1} = T$, where $\mathfrak{R}_{90}$ denotes a 90° rotation operator ($\mathfrak{R}_{90} \vec{e}_x = \vec{e}_y$, $\mathfrak{R}_{90} \vec{e}_y = -\vec{e}_x$). Then, Eq. (1) can be rewritten as

$$\begin{Bmatrix} t_{xx} & t_{xy} \\ t_{yx} & t_{yy} \end{Bmatrix} \times \begin{pmatrix} 0 \\ E_x^{in} \end{pmatrix} = \begin{pmatrix} -E_y^{out} \\ E_x^{out} \end{pmatrix}. \tag{2}$$

According to Eqs. (1) and (2), we obtain $E_y^{out} = t_{yx} \times E_x^{in} = -t_{xy} \times E_x^{in}$, i.e., $t_{yx} = -t_{xy}$. When this bilayered structure has mirror symmetry in z-direction, i.e., no chirality in wave propagation direction, transmission matrix T should satisfy the relation $\vec{T}_{ij} = \overleftarrow{T}_{ij}$ ($\vec{T}$ and $\overleftarrow{T}$ denote the transmission matrices for the two opposite directions of propagation; i, j = x or y). Meanwhile, the reciprocity [14] requires $\overleftarrow{T}_{ij} = \vec{T}_{ji}$. Thus we obtain $\vec{T}_{ij} = \vec{T}_{ji}$, which means $t_{xy} = t_{yx}$. Under this situation, both $t_{xy}$ and $t_{yx}$ should be zero. That means no cross-polarization conversion will occur in the four-fold rotation symmetric structure when the mirror symmetry in propagation direction appears. To confirm the conclusion, the inset of Fig. 2(a) presents the numerical results for the case when the upper and bottom patterns of the bilayered structure are identical, i.e., the structure has mirror symmetry in z direction. In this case, only a small amount (< -25 dB) of cross-polarized transmission is obtained in the whole frequency range, which is due to e.g. the scattering of the metallic elements as the homogenization is an approximation.

Next, we study the polarization state of the transmitted wave at different frequencies to get an intuitive picture of the polarization change by this bilayered CMM. As both co-polarized and cross-polarized transmissions exist, the electric field of the transmitted wave can be written as

$$\boldsymbol{E}^{out}(\boldsymbol{r},t) = \left(|\boldsymbol{E}_x^{out}|\cos(kz-\omega t)\ |\boldsymbol{E}_y^{out}|\cos(kz-\omega t+\phi)\right),$$

where $\phi$ is the phase difference between the co- and cross-polarized transmissions. Accordingly, the axial ratio $20\log_{10}\left(|\boldsymbol{E}^{out}|_{\max}/|\boldsymbol{E}^{out}|_{\min}\right)$ as a function of frequency is calculated and shown in Fig. 2(b). It is worth noting that all the transmitted waves are almost linearly polarized with the axial ratio better than 40 dB. This means pure optical activity, i.e., polarization azimuth rotation with no perceptible change of ellipticity, is achieved in the whole frequency range. As an example, Fig. 2(d) shows the polarization states of the transmitted wave at six specific frequencies. At low frequency ω = 10.0 GHz, the transmission is relative low [see Fig. 2(d)] and the polarization is slightly off the x axis with an azimuth of 20° [see Fig. 2(c)], which is defined as [15]

$$\gamma = (1/2)\arctan\left[\frac{2\operatorname{Re}\left(\boldsymbol{E}_x^{out}\cdot\boldsymbol{E}_y^{out*}\right)}{\left|\boldsymbol{E}_x^{out}\right|^2-\left|\boldsymbol{E}_y^{out}\right|^2}\right],$$

indicating the angle between the major polarization axis and the x axis [the direction of the incident electric field; see Fig. 1(c)]. As frequency increases, the polarization of the transmitted wave rotates counterclockwise toward y axis, which is accompanied by a rapid increase of the transmission. At frequency ω = 11.48 GHz, the transmission reaches the maximum and a polarization azimuth of 86° is achieved simultaneously [see Fig. 2(c)]. Then the transmission is gradually reduced, while polarization azimuth $\gamma$ continuously goes up to 90°. It follows form Fig. 2(d) that the transmitted waves are nearly polarized along y axis in the frequency range 11.66 ~ 12.10 GHz. With a further increase of frequency, the polarization rotates back toward x axis as shown in Fig. 2(d). The above results confirm the fact that a 90° polarization rotation is achieved by this bilayered CMM. Notably, the optical activity of this bilayered CMM is substantial, considering that the thickness of the present structure is only about 1/30 wavelength around the working frequency. In terms of rotation per material thickness of one wavelength, the optical activity of the present CMM is about 2700°/λ, which is much stronger than that obtained in a chiral system based on electric dipole coupling (250°/λ in Ref. [9]), and that in a metal-on-dielectric chiral system (6°/λ in Ref. [10]). To the best of our knowledge, this is the largest optical activity that can be found in literature.

To get physical insight into the mechanism of this strong optical activity, the surface current distributions of this bilayered CMM at resonance is shown in Fig. 3(a) and (b). Obviously, the proposed bilayered structure [see Fig. 1(b)] can be regarded as a combination of four cut-wire pairs (on the two layers) along the four sides of a square. It is well-known that the magnetic field, which is oriented perpendicular to the plane of a cut-wire pair (on the two layers), causes the excitation of a magnetic dipole with anti-parallel currents in the two cut-wires. As shown in Fig. 3(a) and (b), the magnetic dipoles on both top and down sides are directly excited by the incident wave, whose H field is along y-direction. Since the magnetic dipoles in the four identical cut-wire pairs are exactly the same, they can be strongly coupled to each other. Due to this transverse magnetic dipole coupling, significant anti-parallel currents on left and right sides [see Fig. 3(a) and (b)] are also obtained, which results in the cross-polarized transmission (H field is along x-direction) and the strong optical activity in the present bilayered CMM. For comparison, Fig. 3(c) and (d) show the case when the bilayered structure has mirror symmetry in the propagation direction. Almost no current distribution could be observed in the cut-wire pairs on the left or right side. This is because the magnetic dipole coupling, as well as the cross-polarized conversion, is forbidden by the symmetry of the structure as demonstrated above.

Finally, we want to explore the isotropic behavior of this 90° polarization rotator when the polarization azimuth of the incident wave changes. In our experiment, the two horns are fixed, while the sample in the middle is rotated up to 45° about z axis with a step of 5° [see Fig. 2(c)]. The measured cross-polarized transmission spectra as a function of rotation angle $\theta$ are shown in Fig. 4. Due to the four-fold rotational symmetry of the structure, the cross-polarized transmission is not obviously changed with the variation in $\theta$ [see Fig 4(a)], considering the experimental error in measurements. This means the proposed 90° polarization rotator can work efficiently for any polarization azimuth. This is very much different from the case in Ref. [12], where the polarization rotation is dependent on the polarization azimuth of the incident wave.

In summary, we have proposed a new bilayered CMM with enantiomeric patterns of four-fold rotational symmetry to realize a 90° polarization rotator, which exhibits transversely isotropic character. The transverse magnetic dipole coupling has been demonstrated to cause giant optical activity in the proposed CMM. Both the numerical simulation and experimental measurement have been given at microwave frequencies and they agree well with each other. Due to its thin thickness and high polarization conversion efficiency, the present bilayered CMM has giant optical activity (largest in literature) and the associated 90° polarization rotator may have potential applications in e.g. telecommunication.

The authors would like to acknowledge Dr. Hongshen Chen for his help in experimental measurements. We also acknowledge the partial support of National Basic Research program (973 Program) of China (No.2004CB719800), a NSFC (No.60890070) and a Swedish Research Council (VR) grant on metamaterials.

* sailing@kth.se

**Figure captions:**

FIG. 1. (a) Unit cell of the present bilayered CMM for transmission calculation. (b) The top view of the fabricated microwave-scale sample. (c) Schematic picture of the experimental setup, the incident wave is emitted from the source and received by the detector. Axes indicate the propagation direction and polarization of the incident plane wave.

FIG. 2. (Color online) (a) Co-polarized (red) and cross-polarized (blue) transmission of electric field through the present bilayered CMM by both simulation (solid lines) and experimental measurement (dotted lines). (b) Axial ratio and (c) polarization azimuth of the transmitted wave as the frequency varies. (d) Polarization states at six different frequencies.

FIG. 3. (Color online) The surface current distributions on the (a) upper and (b) bottom patterns of the present bilayered CMM at resonance. (c), (d) shows the case when the bilayered structure has the same pattern on the upper and bottom sides. The black arrows denote the surface current directions.

FIG. 4. (Color online) The measured cross-polarized transmission spectra of the present bilayered CMM as a function of the rotation angle $\theta$.

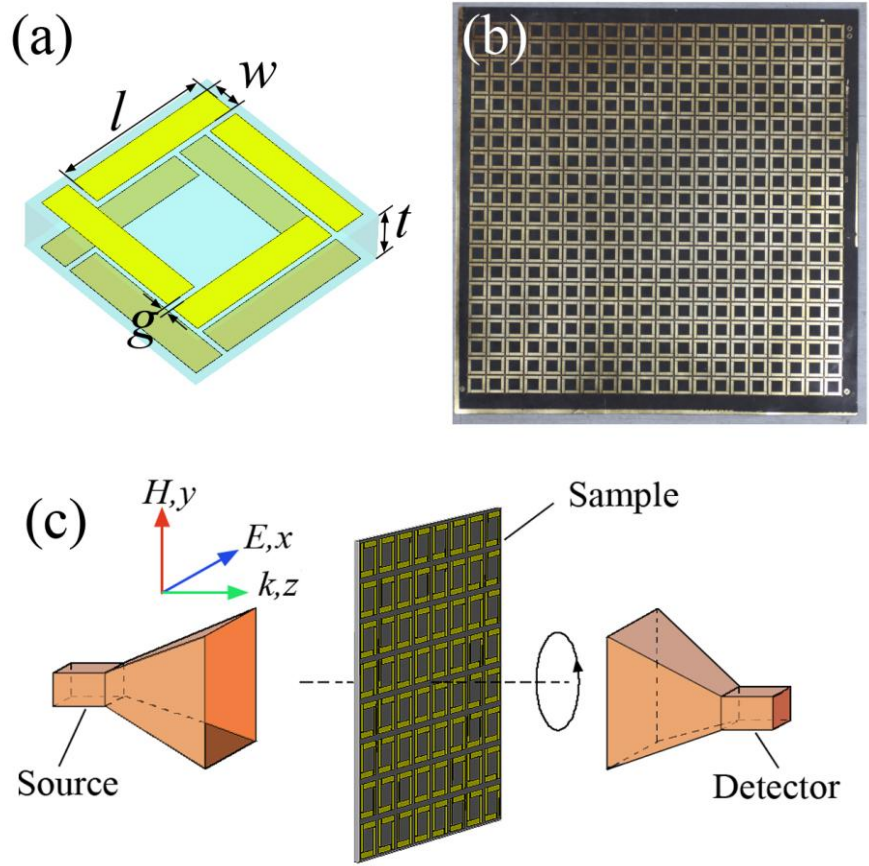

FIG. 1. (a) Unit cell of the present bilayered CMM for transmission calculation. (b) The top view of the fabricated microwave-scale sample. (c) Schematic picture of the experimental setup, the incident wave is emitted from the source and received by the detector. Axes indicate the propagation direction and polarization of the incident plane wave.

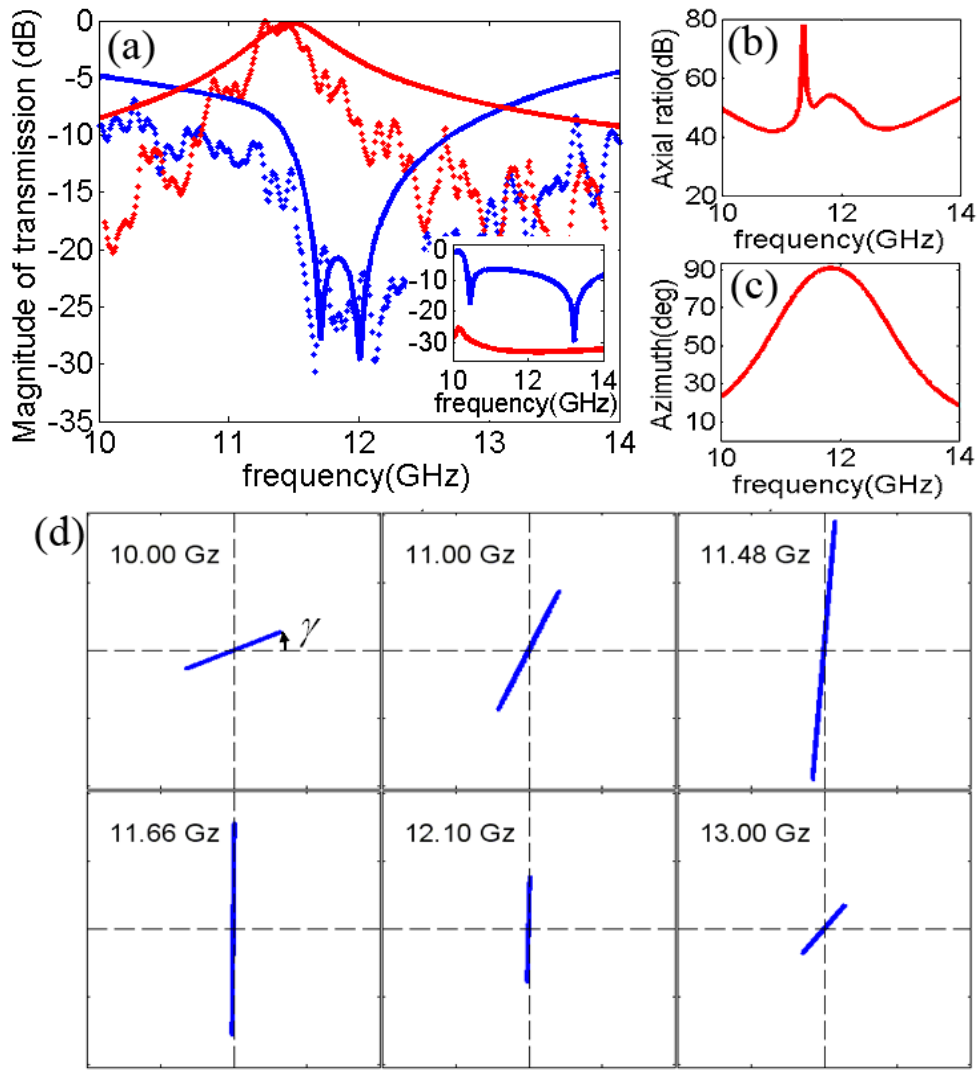

FIG. 2. (Color online) (a) Co-polarized (red) and cross-polarized (blue) transmission of electric field through the present bilayered CMM by both simulation (solid lines) and experimental measurement (dotted lines). (b) Axial ratio and (c) polarization azimuth of the transmitted wave as the frequency varies. (d) Polarization states at six different frequencies.

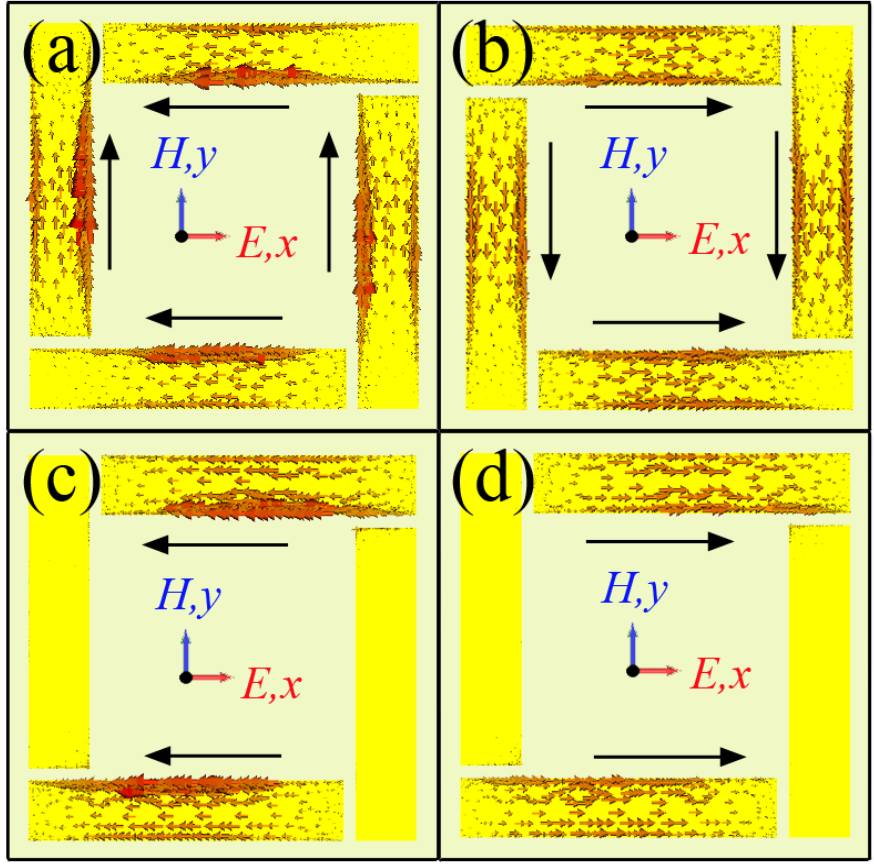

FIG. 3. (Color online) The surface current distributions on the (a) upper and (b) bottom patterns of the present bilayered CMM at resonance. (c), (d) shows the case when the bilayered structure has the same pattern on the upper and bottom sides. The black arrows denote the surface current directions.

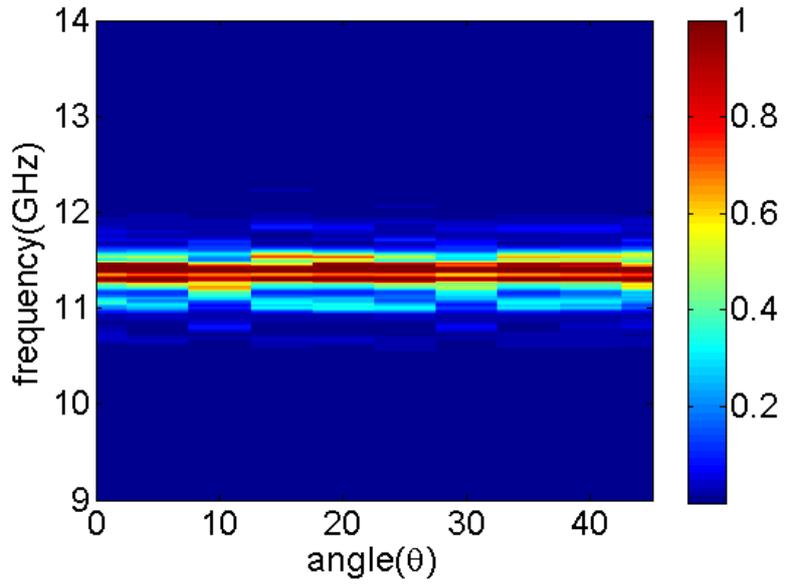

FIG. 4. (Color online) The measured cross-polarized transmission spectra of the present bilayered CMM as a function of the rotation angle θ.